\documentclass[12pt]{article}
\usepackage{a4wide,makeidx,color,latexsym}			
\usepackage{amsmath,amsfonts,verbatim,graphicx,float,amssymb,url,subfigure,epsfig,appendix}
\usepackage{xspace}
\usepackage{a4wide,makeidx,color,latexsym}			
\usepackage{amsmath,amsfonts,verbatim,graphicx,float,amssymb,url,subfigure,epsfig,appendix}

\newcommand{\pp}{pp\xspace}
\newcommand{\sppt}[1]{\ensuremath{\sqrt{s} = #1}\xspace}
\newcommand{\snnnotext}[1]{\ensuremath{\sqrt{s_{\rm NN}} = #1}\xspace}
\newcommand{\pbpb}{Pb-Pb\xspace}
\newcommand{\dedx}{\ensuremath{\text{d}E/\text{d}x}\xspace}
\def\beq{\begin{equation}}
\def\eeq#1{\label{#1}\end{equation}}
\def\eeqn{\end{equation}}

\def\beqa{\begin{eqnarray}}
\def\eeqa#1{\label{#1}\end{eqnarray}}
\def\eeqan{\end{eqnarray}}

\let\bar=\overbar

\def\Dslash{\not{\hbox{\kern-4pt $D$}}}
\def\dslash{\not{\hbox{\kern-2pt $\del$}}}

\def\msb{{\bar{\ssstyle M \kern -1pt S}}}

\def\Title#1{\begin{center} {\Large {\bf #1} } \end{center}}

\begin{document}

\Title{Identified particle $p_{\rm T}$ spectra and particle contents  in \pp collisions measured with ALICE at the LHC}

\bigskip\bigskip


\begin{raggedright}  

{\it Antonio Ortiz Velasquez (on behalf of the ALICE Collaboration)\index{Ortiz, A.}\\
Div. of Particle Physics\\
Lund University\\
SE-221 00 Lund, SWEDEN}
\bigskip\bigskip
\end{raggedright}

\section{Introduction}

ALICE is a general-purpose heavy-ion experiment at the LHC. The design is optimized for reconstruction and particle identification (PID) in a wide range of transverse momentum ($p_{\rm T}$) \cite{ref:0,ref:0b}.  Since 2009 it has collected data from \pp collisions at \sppt{0.9}, 2.76, 7 and 8 TeV.

The main focus of ALICE is the study of \pbpb collisions. Measurements of relevant observables in \pp collisions constitute a baseline for the interpretation of the results in nucleus-nucleus collisions. Moreover, the ALICE capabilities allow also to study some important aspects of \pp physics. For example, at LHC energies the bulk of the particles produced at mid-rapidity have low transverse momentum ($<$ 2 GeV/$c$). Since perturbative Quantum Chromo-Dynamics (p-QCD) is not applicable, the particle production is modeled following phenomenological approaches. Hence, the measurement of identified particle production at low $p_{\rm T}$ is a valuable input for models of hadronization process. For higher momenta ($>$ 7 GeV/$c$), $p_{\rm T}$ spectra of identified particles provide important information to constrain the fragmentation functions (FFs) at large $z$ (momentum fraction of the hadron relative to the parent parton).

The relevant detectors for PID which were used to produce the present results are: Inner Tracking System (ITS), Time Projection Chamber (TPC) and Time Of Flight detector (TOF) \cite{ref:1}, they are located in the central barrel of ALICE inside a large solenoidal magnet providing a uniform 0.5 T field~\cite{ref:0,ref:0b}. 

The ITS is composed of six cylindrical layers of silicon detectors. The two innermost layers are pixel detectors (SPD), followed by two layers of drift detectors (SDD) and two layers of double-sided silicon strip detectors (SSD). SDD and SSD provide measurements of the specific energy loss \dedx for charged particles.

The TPC is the main tracking device. It is a large cylindrical drift detector with a central membrane maintained at -100 kV and two readout planes at the end-caps composed of 72 multi-wire proportional chambers.  The active volume is limited to $85 < r < 247$ cm and $−250 < z < 250$ cm in the radial and longitudinal directions, respectively. With this detector we can measure the specific energy loss \dedx relying on a sample of up to 159 points per charged track. Charged pions, kaons and (anti)protons are well separated at low $p_{\rm T}$ ($<1$ GeV/$c$). The identification can be extended up to 20 GeV/$c$ using a statistical approach based on the relativistic rise where the separation of particles with different masses is almost constant.

The TOF detector consists of 18 azimuthal sectors, each containing 91 Multi-gap Resistive Plate Chambers (MRPCs) distributed in five gas-tight modules. Particles are identified by measuring their momentum and velocity simultaneously.

The strange and multi-strange particles: ${\rm K}_{S}^{0}$, $\phi$, $\Lambda$($\bar{\Lambda}$), $\Xi^{-}$($\bar{\Xi}^{+}$) and $\Omega^{-}$($\bar{\Omega}^{+}$) are identified from their weak decay topologies, more details can be found in \cite{ref:2, ref:3}.

\section{Results}

\begin{figure}[htbp]
\begin{center}
\mbox{
\epsfig{file=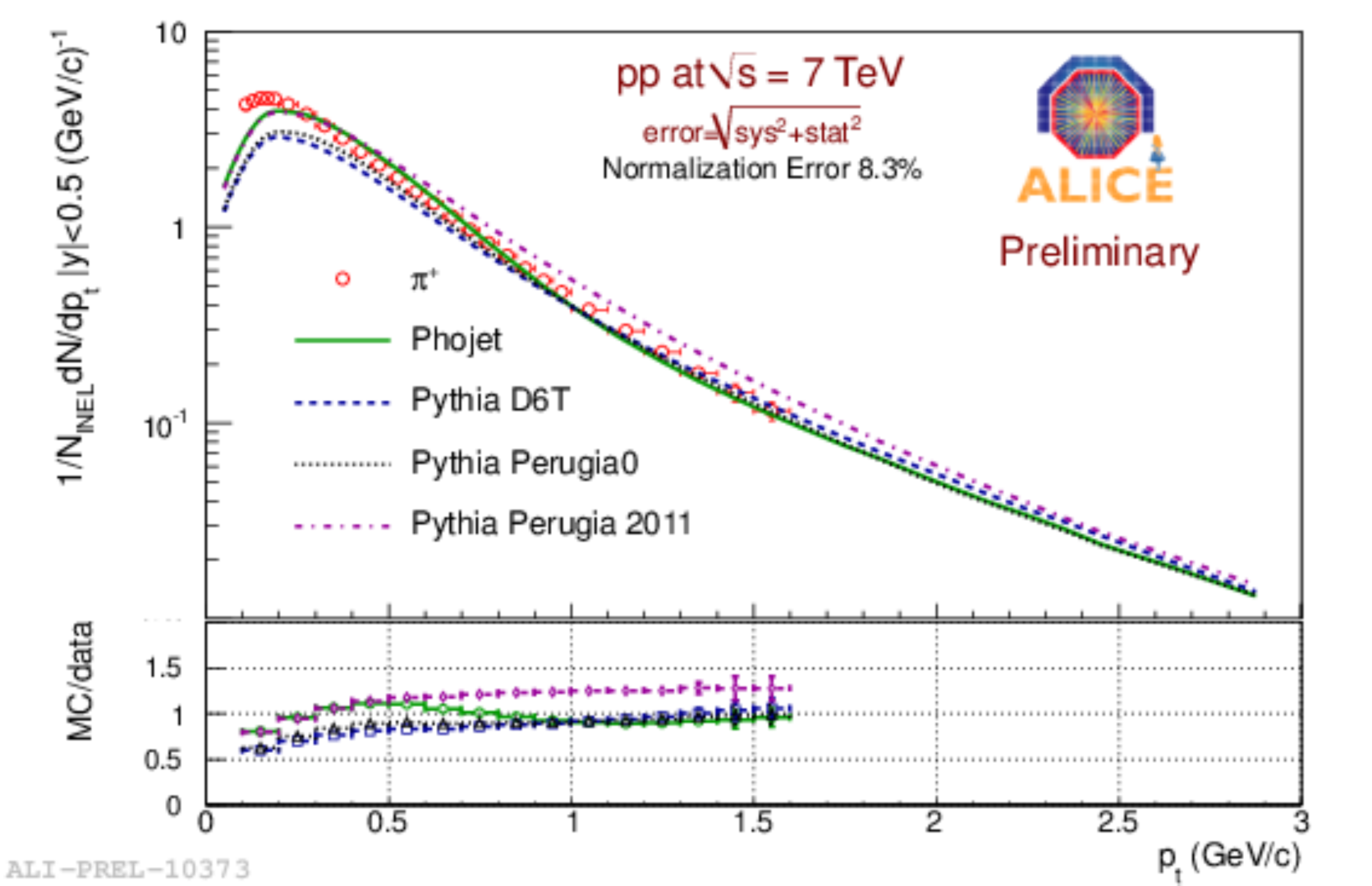,width=0.335\columnwidth}
\epsfig{file=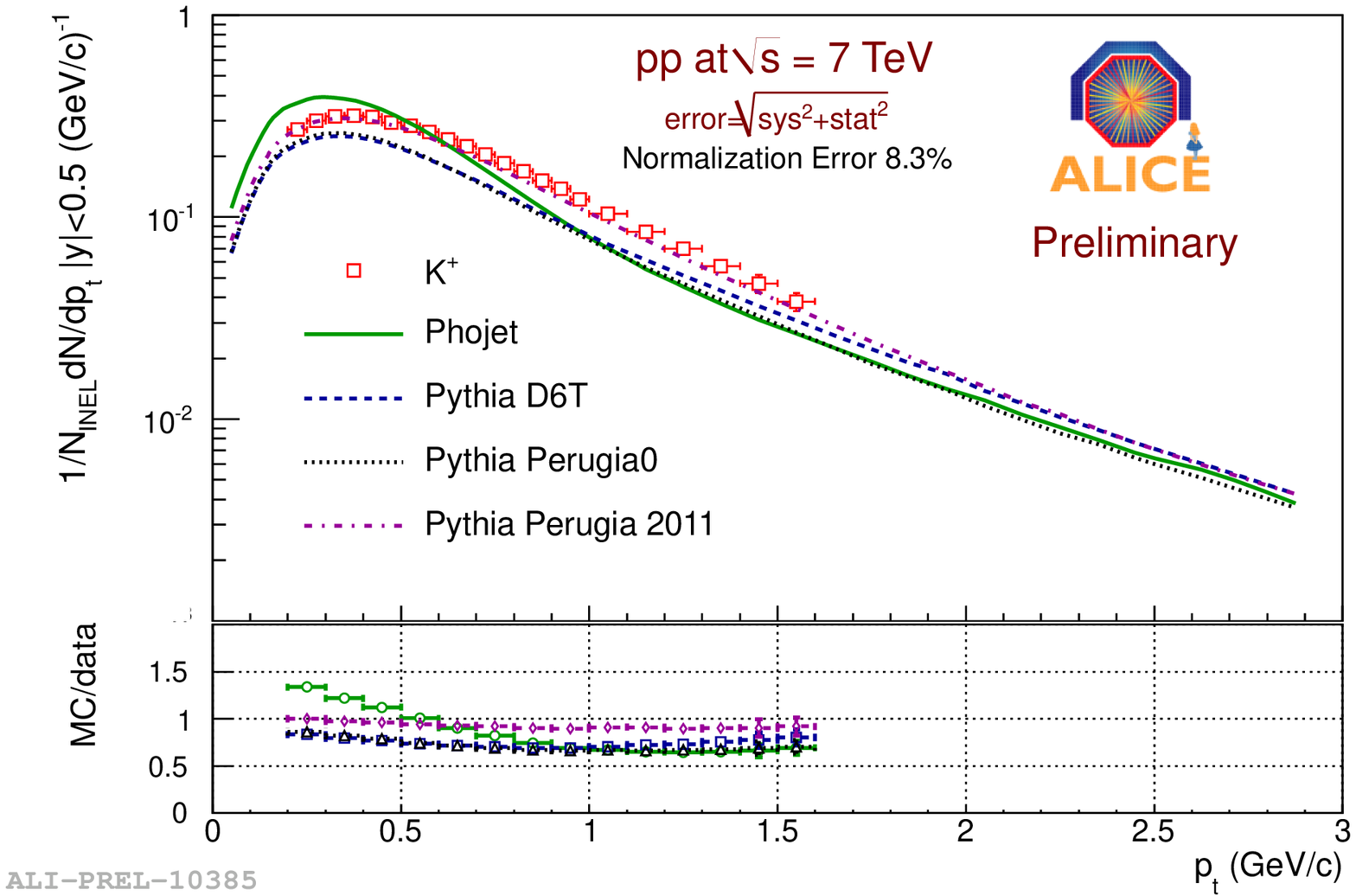,width=0.335\columnwidth}
\epsfig{file=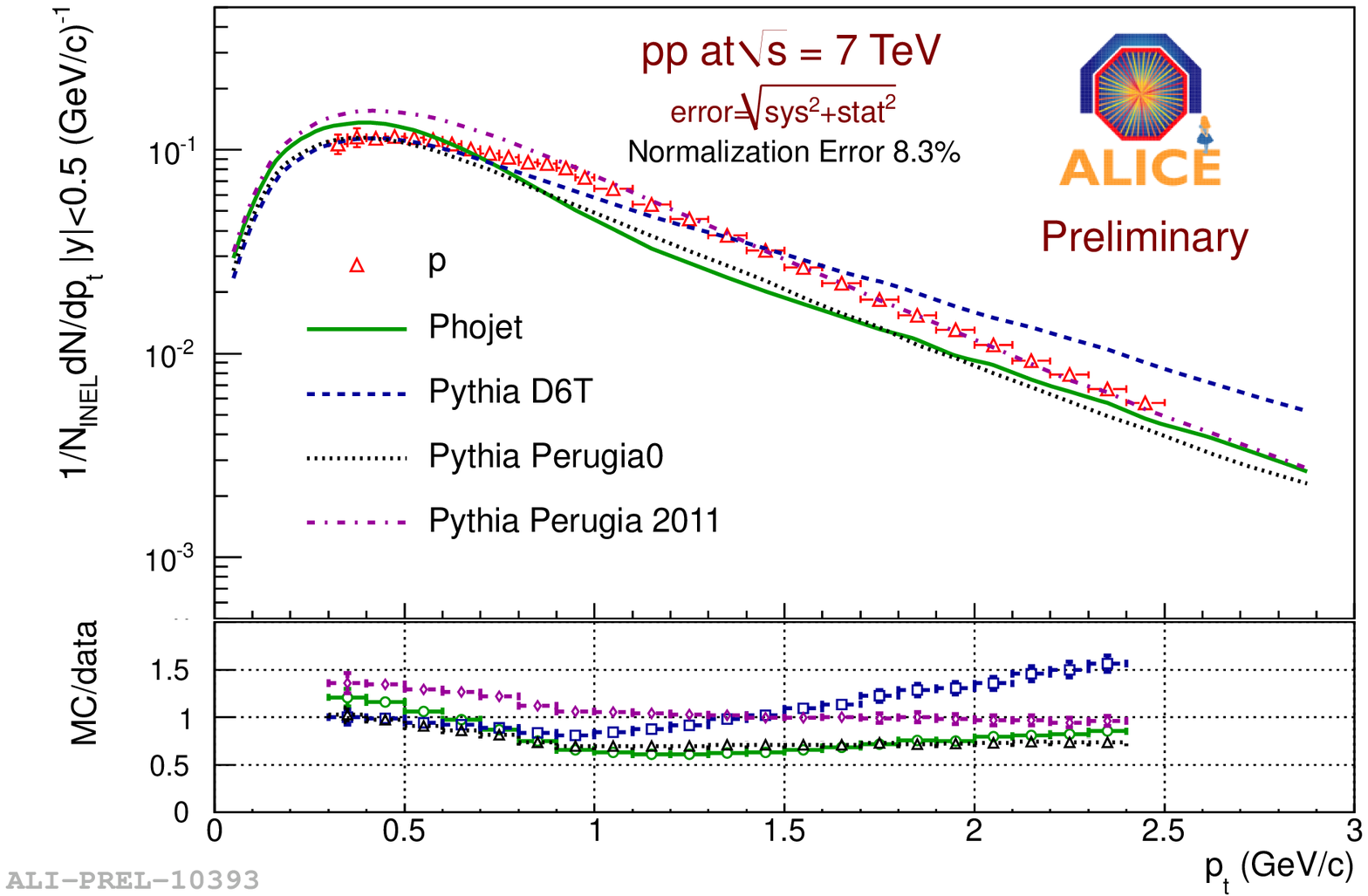,width=0.335\columnwidth}
}
\caption{Comparison of the measured $\pi^{+}$ (left), ${\rm K}^{+}$ (middle) and $\rm p$ (right) $p_{\rm T}$ spectra  in \pp collisions at \sppt{7} TeV with predictions from Phojet and Pythia-6.}
\label{fig:1}
\end{center}
\end{figure}

Figure~\ref{fig:1} shows the  $\pi^{+}$, ${\rm K}^{+}$ and $\rm p$ transverse momentum spectra measured in \pp collisions at \sppt{7} TeV. ALICE results are compared with predictions from Phojet~\cite{ref:4} and Pythia-6~\cite{ref:5} (tunes: D6T, Perugia-0 and Perugia-2011). We observe that the models cannot describe the three yields simultaneously, {\it e.g.} Perugia-2011 only describes quite well the kaon yield. We obtain similar results if we consider their antiparticles, actually the $\overline{\rm p}/{\rm p}$ ratio is independent of both rapidity and transverse momentum \cite{ref:11}.

\begin{figure}[htbp]
\begin{center}
\mbox{
\epsfig{file=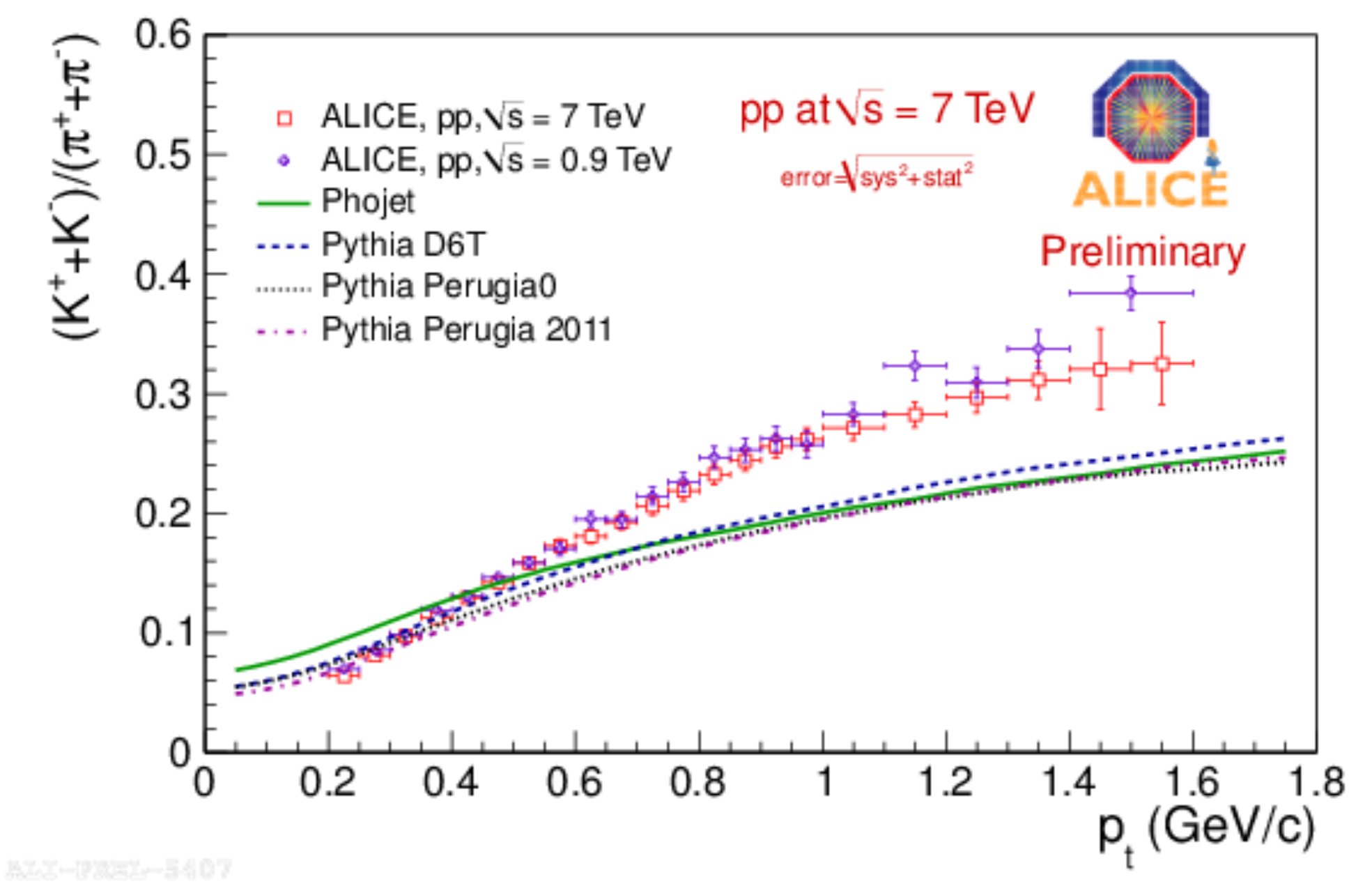,width=0.47\columnwidth}
\epsfig{file=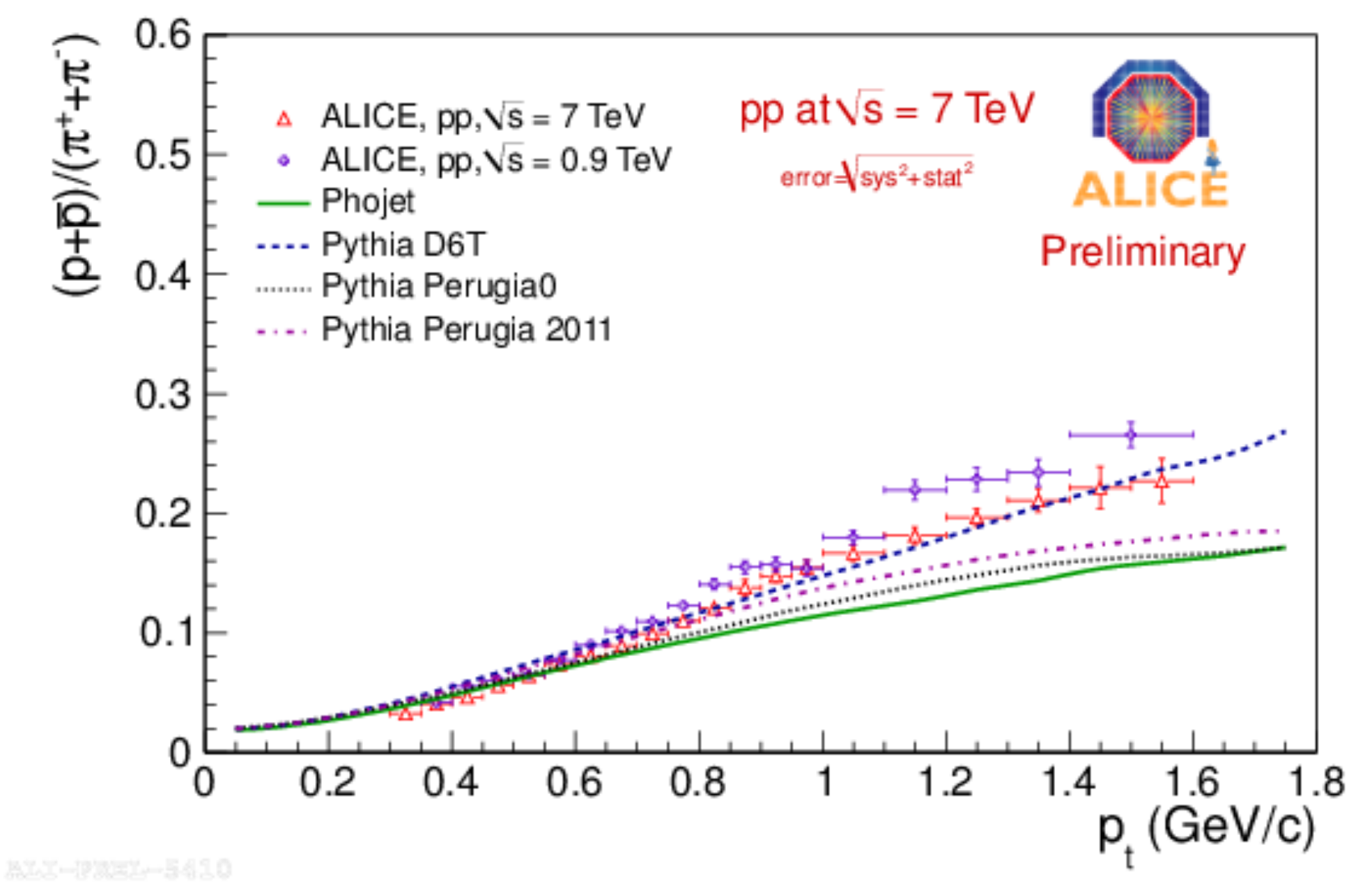,width=0.47\columnwidth}
}
\caption{$({\rm K}^{+}+ {\rm K}^{-})$/$(\pi^{+}+\pi^{-})$ and $({\rm p + \bar{p}})$/$(\pi^{+}+\pi^{-})$ ratios as a function of $p_{\rm T}$ measured in \pp collisions at \sppt{0.9} and 7 TeV. ALICE data are compared with predictions from Phojet and Pythia-6.}
\label{fig:2}
\end{center}
\end{figure}

The ratios $({\rm K}^{+}+ {\rm K}^{-})$/$(\pi^{+}+\pi^{-})$ and $({\rm p + \bar{p}})$/$(\pi^{+}+\pi^{-})$ can be used to probe the fraction of strange to non-strange particles and the baryon to meson ratio, respectively. Figure~\ref{fig:2} shows both ratios as a function of $p_{\rm T}$ for \pp collisions at \sppt{0.9} and 7 TeV. Within systematic and statistical uncertainties, both ratios seem to be energy independent, moreover, from \cite{ref:1} one can see that K/$\pi$ is constant from \sppt{0.2} up to 7 TeV. Except from Pythia-6 tune D6T which describes well $\rm p$/$\pi$, the event generators investigated here do not reproduce the data.

\begin{figure}[htbp]
\begin{center}
\mbox{
\epsfig{file=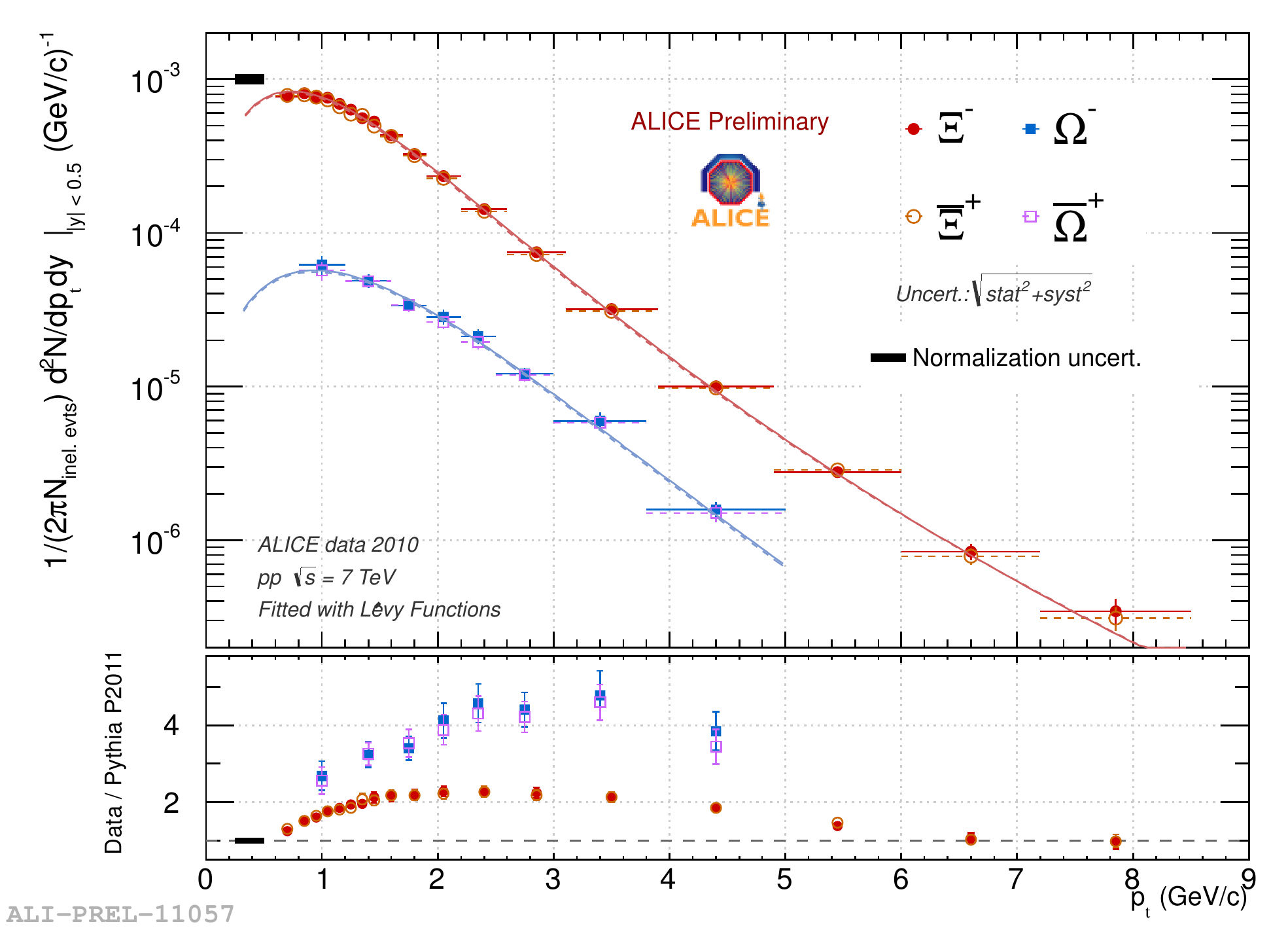,height=2.42in,width=0.45\columnwidth}
\epsfig{file=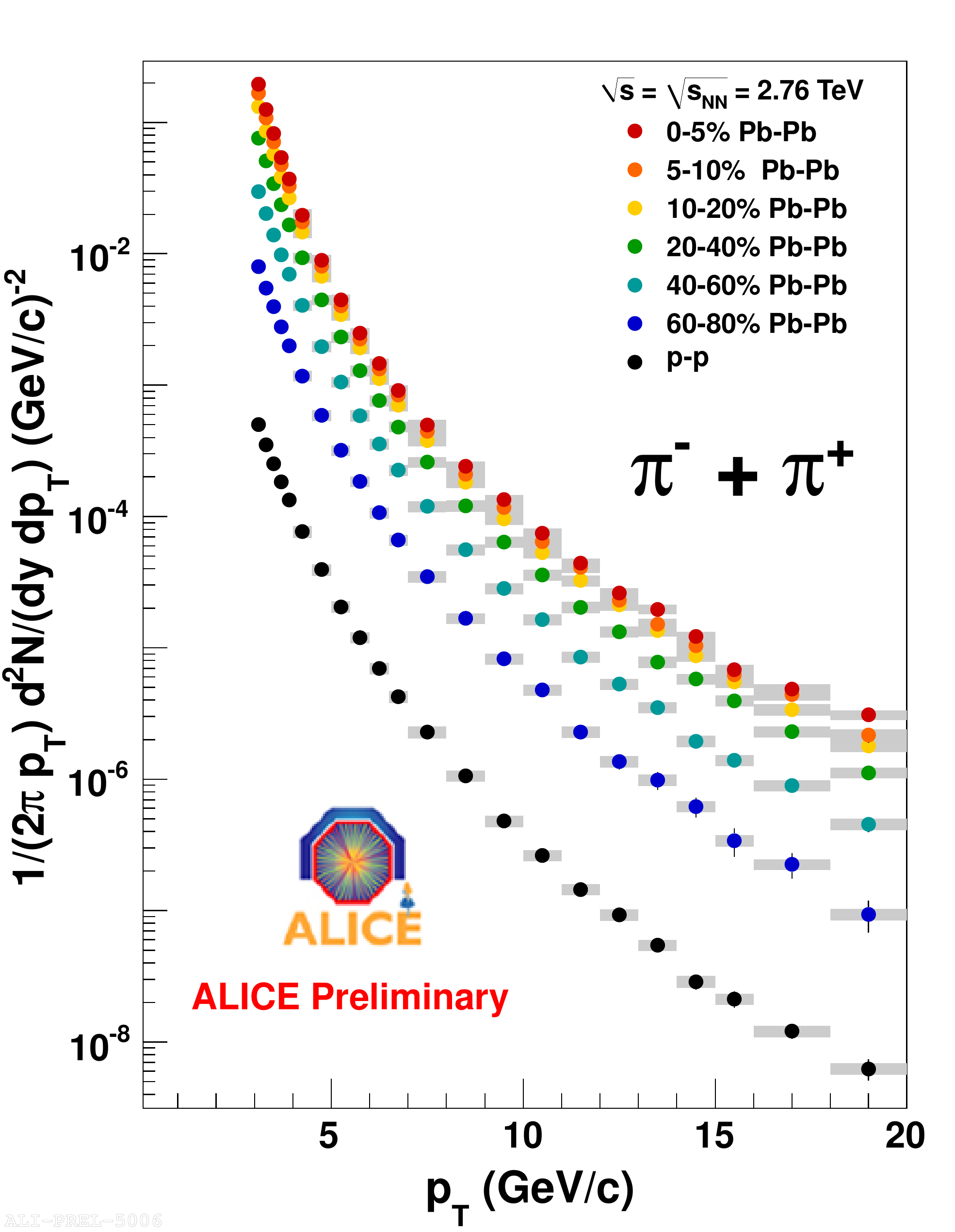,height=2.5in,width=2.0in}
}
\caption{ (Left) $\Xi^{-}$($\bar{\Xi}^{+}$) and $\Omega^{-}$($\bar{\Omega}^{+}$) measured baryon spectra in \pp collisions at \sppt{7} TeV superimposed with Tsallis fits~\cite{ref:3}. The bottom panel shows a comparison with Pythia-6 tune Perugia-2011. (Right) Measured high-$p_{\rm T}$ charged pion spectra from \pbpb and \pp collisions at \snnnotext{2.76} TeV. For heavy-ions, different centrality classes are shown.}
\label{fig:3}
\end{center}
\end{figure}

The agreement between data and event generators is even worse for predictions of strange baryon production~\cite{ref:2,ref:3}. For instance, at \sppt{0.9} TeV the production of $\Lambda$ is underestimated up to a factor $\sim$ 5. For multi-strange baryons, Figure \ref{fig:3} shows the measured $\Xi^{-}$ ($\bar{\Xi}^{+}$) and $\Omega^{-}$ ($\bar{\Omega}^{+}$) spectra for \sppt{7} TeV data. The results are compared with Perugia-2011. The agreement between data and theory is better for $\Xi^{-}$ ($\bar{\Xi}^{+}$) and specially at high transverse momentum ($>5$ GeV/$c$). However, for lower $p_{\rm T}$ (1-4 GeV/$c$) theory underestimates the production by a factor $\sim2$. On the other hand, the production of $\Xi^{-}$ ($\bar{\Xi}^{+}$) is underestimated by a factor larger than 3.

The measurement of TPC-\dedx on the relativistic rise allows to extend the identification of $\pi^{\pm}$/${\rm K}^{\pm}$/${\rm p}$($\bar{{\rm p}}$) up to $\sim 20$ GeV/$c$. Figure \ref{fig:3} also shows the charged pion spectra measured in \pp and \pbpb collisions at \snnnotext{2.76} TeV. The spectrum for \pp collisions seems harder than the ones for central \pbpb collisions because the medium suppresses the production of high $p_{\rm T}$ particles.



\section{Conclusions}

Transverse momentum spectra for identified particles were measured by ALICE. So far the event generators which were tested do not describe  the various observables simultaneously. That is true even for Perugia-2011 which  was tuned using the early LHC data and optimized to increase the strange hadron production~\cite{ref:8}. This tune describes reasonably well the inclusive charge particle $p_{\rm T}$ spectrum at 7 TeV. However it does not succeed in describing the particle composition, especially the multi-strange baryon production. The aforementioned observations represent a challenge for hadro-production models.

The measured high-$p_{\rm T}$ charged pion production is important, because it provides data to probe the Parton Distribution Functions and Fragmentation Functions at low $x$ and large $z$, respectively. 

\def\Discussion{
\setlength{\parskip}{0.3cm}\setlength{\parindent}{0.0cm}
     \bigskip\bigskip      {\Large {\bf Discussion}} \bigskip}
\def\speaker#1{{\bf #1:}\ }
\def\endDiscussion{}



 
\end{document}